\documentclass[conference]{IEEEtran}

\usepackage{cite}
\usepackage{amsmath,amssymb,amsfonts}
\usepackage{algorithmic}
\usepackage{graphicx}
\usepackage{textcomp}
\usepackage{xcolor}
\usepackage{xurl}
\usepackage[hidelinks]{hyperref}
\usepackage{array}
\usepackage{tabularx}
\usepackage{booktabs}
\usepackage{braket}
\usepackage{lipsum}
\usepackage{booktabs,array}
\usepackage[a4paper, total={184mm,239mm}]{geometry}
\usepackage{subcaption}
\def\BibTeX{{\rm B\kern-.05em{\sc i\kern-.025em b}\kern-.08em
    T\kern-.1667em\lower.7ex\hbox{E}\kern-.125emX}}

\newcommand{\qsink}{\texttt{qSink}}
\newcommand{\qsinkc}{\texttt{qSink-compiler}}
\newcommand{\qsinkf}{\texttt{qSink-fetcher}}
\newcommand{\idletime}{$79\%$}
\newcommand{\speedup}{$1.18\times$}
\newcommand{\idleovertotal}{$55\%$}
\newcommand{\combspeedup}{$1.32\times$}
\newcommand{\swiper}{speculative window decoding}

\begin{document}

\title{qSink: Hiding QEC Decoder Latency through Conditional-Correction Sinking\\
\thanks{Identify applicable funding agency here. If none, delete this.}
}

\author{
\IEEEauthorblockN{
Alessio Cicero\textsuperscript{1},
Luigi Altamura\textsuperscript{1},
Mats Granath\textsuperscript{2},
and Pedro Trancoso\textsuperscript{1}
}
\IEEEauthorblockA{
\textsuperscript{1}\textit{Chalmers University of Technology and University of Gothenburg},
Gothenburg, Sweden\\
\textsuperscript{2}\textit{University of Gothenburg},
Gothenburg, Sweden\\
\{ciceroa, altamura, ppedro\}@chalmers.se,
mats.granath@physics.gu.se
}
}

\maketitle

\begin{abstract}
Parallel window decoding is a leading approach to scalable quantum error correction because it increases decoder throughput by processing decoding windows concurrently. However, dependencies between decoding windows can delay the decoder feedback required for measurement-dependent execution. In fault-tolerant Clifford+$T$ circuits, for example, the decoded outcome of a $T$-gate teleportation determines whether a conditional $S$ correction is required, forcing dependent execution to wait until this result becomes available. Across our evaluation of QASMBench circuits spanning diverse algorithms, gate compositions, and circuit scales, these decoding-induced stalls account for up to \idleovertotal{} of total execution time.
\par
To address this bottleneck, we propose \qsink{}, a scheduling methodology that hides decoding latency by moving conditional corrections past compatible subsequent Clifford+$T$ operations, allowing these operations to execute earlier while decoding proceeds. We integrate \qsink{} into the Lattice Surgery Compiler to build \qsinkc, a quantum-error-correction-aware compiler that applies these transformations and annotates the resulting schedule with decoder-dependency metadata. We also implement \qsinkf, a lightweight FPGA-based schedule preprocessor that consumes \qsinkc-generated schedules and uses their dependency metadata together with decoder feedback to correctly resolve conditional operations at runtime.
\par
Across the evaluated workloads, \qsink{} reduces decoding-induced idle time by up to \idletime{} and achieves an end-to-end speedup of up to \speedup. Its idle-time reduction is comparable to speculative window decoding, the most closely related state-of-the-art decoder-side optimization. Because \qsink{} delays when decoder feedback is required while speculative window decoding produces that feedback earlier, combining the two achieves up to \combspeedup{} end-to-end speedup.



\end{abstract}


\section{Introduction}
Quantum error correction~(QEC) is essential for fault-tolerant quantum computation, as physical-qubit error rates are too high to support long-running quantum algorithms directly~\cite{fowler2012surface}. However, decoding the continuously generated syndrome data becomes increasingly challenging as code distance and system size grow. Two approaches have been proposed to address this scalability problem: sliding window decoding~\cite{dennis2002topological-sliding-window} and parallel window decoding~\cite{skoric2023parallel}. Parallel window decoding provides greater scalability by processing multiple decoding graphs concurrently, increasing decoding throughput and relaxing the latency requirements placed on the decoder.\par

\begin{figure}
    \centering
    \includegraphics[width=0.7\linewidth]{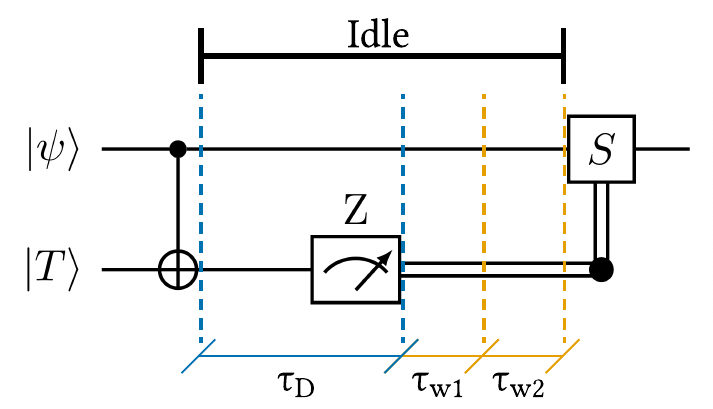}
    \caption{Fault-tolerant $T$-gate teleportation and decoding-induced idle time. The decoded $Z$-measurement outcome determines whether the conditional $S$ correction is required. $\tau_D$ denotes the buffer-region generation time~\cite{skoric2023parallel}, while $\tau_{w1}$ and $\tau_{w2}$ denote the latencies of the two dependent decoding windows.}
    \label{fig:t-gate}
\end{figure}

While parallel window decoding addresses decoder throughput, dependencies between decoding windows can delay the feedback required for measurement-dependent execution. Fault-tolerant circuits are commonly expressed using the Clifford+$T$ gate set~\cite{litinski2019game-of-surface-codes}, where $T$ gates introduce measurement-dependent classical control. As shown in Figure~\ref{fig:t-gate}, the measurement must be decoded before the controller can determine whether a conditional $S$ correction is required. While Clifford operations can generally proceed without immediately resolving decoder output, a dependent operation following a $T$ gate may therefore stall until the required decoder feedback becomes available. Across our evaluated QASMBench~\cite{li2023qasmbench} workloads, these decoding-induced stalls account for up to \idleovertotal{} of total execution time. Thus, even when parallel decoding provides sufficient throughput to prevent a backlog, delayed decoder feedback can still stall program execution.\par


Prior work reduces this waiting time from the decoder side. In particular, speculative window decoding~\cite{viszlai2025swiper} predicts dependencies between adjacent decoding windows, allowing the dependent window to begin decoding speculatively during $\tau_{w1}$. As shown in Figure~\ref{fig:t-gate}, $\tau_{w1}$ and $\tau_{w2}$ represent the two dependent decoding-window latencies: when the speculation is correct, the required feedback is available after $\tau_{w1}$; otherwise, decoding extends into $\tau_{w2}$ to resolve the dependency.\par
\qsink{} instead acts on the program side by moving the conditional $S/S^\dagger$ correction of a $T/T^\dagger$ gate past compatible subsequent operations. This allows post-$T$ execution to continue while decoding proceeds, delaying when the decoder result becomes necessary and enabling overlap with the full feedback interval, including $\tau_D$, $\tau_{w1}$, and $\tau_{w2}$.\par

We realize \qsink{} in \qsinkc{}, a QEC-aware extension of the Lattice Surgery Compiler~(\texttt{liblsqecc}~\cite{watkins2024high-liblsqecc}) flow. \qsinkc{} identifies post-$T$ contexts matching the \qsink{} transformations and moves compatible operations before the pending conditional correction. Conditional corrections can commute directly through $S/S^\dagger$ gates, be transformed across $H\rightarrow S$ sequences into simpler conditional operations, or be moved past $Z$-basis measurements. The rewritten program is then scheduled by the unmodified lattice-surgery scheduler, after which \qsinkc{} annotates decoder-dependent corrections with their controlling measurement outcomes. Importantly, \qsinkc{} leaves both the underlying decoder and lattice-surgery scheduler unchanged, making \qsink{} complementary to the body of work optimizing decoder algorithms and hardware~\cite{barber2025real-synthetic-decoder-data, alavisamani2024promatch,wu2025micro,cicero2026low}.\par

Because \qsink{} moves operations across decoder-dependent corrections, these dependencies must remain associated with the measurement outcomes in the transformed schedule. We therefore introduce \qsinkf{}, a lightweight FPGA-based schedule preprocessor placed between the decoder, schedule memory, and quantum controller. \qsinkf{} consumes the schedules generated by \qsinkc{} and uses their dependency metadata together with decoder feedback to determine whether each conditional correction should be retained or removed before forwarding the updated schedule to the quantum controller. All scheduling decisions remain static: \qsinkf{} neither recompiles nor reschedules the program, but resolves the pre-generated conditional corrections at runtime with minimal additional hardware.\par

Our evaluation shows that this compiler approach can hide a substantial fraction of decoding-induced idle time. Across code distances $d=11$ to $d=31$, \qsink{} reduces decoding-induced idle time by up to approximately \idletime{} and achieves up to approximately \speedup{} end-to-end speedup. Within the evaluated range, the benefit of the two techniques reflects their different scaling: \qsink{} exploits compatible post-$T$ operations, whose duration scales with code distance, while speculative window decoding benefits from the faster growth of decoder-window latency in our model. The two approaches are complementary: \qsink{} delays when decoder feedback is required, while speculative window decoding makes that feedback available earlier. Combining them reaches approximately \combspeedup{} speedup in our code distance sweep. Finally, our FPGA implementation shows that \qsinkf{} uses at most $1.97\%$ of the LUTs required by the reference real-time QEC decoder~\cite{wu2025micro}.\par
Our work makes the following contributions:
\begin{itemize}
\item \textbf{Decoding-induced stall characterization.} We characterize the critical-path operations following $T/T^\dagger$ gates across QASMBench workloads~\cite{li2023qasmbench} and show that decoding-induced feed-forward stalls can account for up to \idleovertotal{} of total execution time. We identify the post-$T$ gate patterns that provide opportunities for hiding this latency.

\item \textbf{\qsinkc.} We introduce \qsinkc{}, a QEC-aware compiler transformation that moves conditional $S/S^\dagger$ corrections across compatible Clifford+$T$ operations. \qsinkc{} rewrites compatible post-$T$ contexts, relies on the unmodified lattice-surgery scheduler to generate the schedule, and annotates decoder-dependent corrections with their controlling measurement outcomes.

\item \textbf{\qsinkf.} We design and implement \qsinkf{}, a lightweight FPGA-based schedule preprocessor that consumes \qsinkc-generated schedules and uses their dependency metadata together with decoder feedback to resolve conditional operations at runtime. Our implementation shows that supporting \qsink{} requires at most $1.97\%$ of the LUTs used by the reference QEC decoder~\cite{wu2025micro}.

\item \textbf{Composition with speculative window decoding.} We show that \qsink{} is complementary to speculative window decoding: \qsink{} delays when decoder feedback is required, while speculative window decoding makes that feedback available earlier. Combining the two achieves up to $1.3\times$ end-to-end speedup.

\end{itemize}
\qsinkc{} is publicly available at \url{https://github.com/mach-research-lab/qsink-compiler}.

\section{Background and Motivation}
\label{sec:bg}
\subsection{Fault-Tolerant Execution and Parallel Decoding}

Quantum error correction~(QEC) protects quantum information by encoding logical qubits into many noisy physical qubits. In surface-code architectures, logical qubits are represented as two-dimensional patches whose evolution in space and time implements fault-tolerant operations through lattice surgery~\cite{horsman2012surface-latticesurgery}. Logical operations are realized through patch-level actions such as merges, splits, deformations, and joint Pauli measurements.
\par
QEC continuously generates syndrome data that must be decoded while the computation executes. Parallel window decoding~\cite{skoric2023parallel} partitions this stream into overlapping decoding windows, each containing a commit region, where corrections are finalized, and buffer regions that preserve information needed across window boundaries. Independent windows can then be processed concurrently, allowing decoding throughput to scale with the available classical resources. 
\par
We treat the decoder operating on each window as a black box and use the code-distance-dependent latencies reported by Barber \emph{et al.}~\cite{barber2025real-synthetic-decoder-data}, shown in Figure~\ref{fig:execution-vs-idle}.



\subsection{$T$-Gate Reaction Time}

Fault-tolerant programs are commonly decomposed into Clifford+$T$ operations~\cite{litinski2019game-of-surface-codes}. Pauli corrections can be tracked classically using a Pauli frame~\cite{chamberland2018fault-pauli-frame}, allowing Clifford execution to proceed without immediately decoding the output. $T/T^\dagger$ gates, however, introduce a feed-forward dependency: the decoded measurement outcome determines whether a conditional $S/S^\dagger$ correction is required. Consequently, subsequent dependent execution may stall until this result becomes available.

As illustrated in Figure~\ref{fig:t-gate}, we model this stall as
\[
\tau = \tau_D + \tau_{w1} + \tau_{w2},
\]
where $\tau_D$ is the time required to generate the syndrome data forming the buffer region, and $\tau_{w1}$ and $\tau_{w2}$ are the latencies of the two dependent windows in the parallel-window decoding sequence. We model the buffer region as $d$ syndrome rounds~\cite{viszlai2025swiper}; assuming a $1\,\mu\mathrm{s}$ syndrome-extraction cycle gives $\tau_D=d\,\mu\mathrm{s}$. Speculative window decoding~\cite{viszlai2025swiper} predicts the dependency between adjacent decoding windows, allowing the second dependent window to begin decoding speculatively during $\tau_{w1}$. When the speculation is correct, the required feedback becomes available after $\tau_{w1}$; otherwise, decoding extends into $\tau_{w2}$ to resolve the dependency. With approximately $90\%$ correct speculation, the expected stall becomes $\tau_D+1.1\tau_w$. Importantly, this optimization reduces the expected waiting time for decoder feedback but does not remove the dependency on the conditional correction.

\subsection{Decoding-induced stall Characterization}

\begin{figure}
    \centering
    \includegraphics[width=0.9\linewidth]{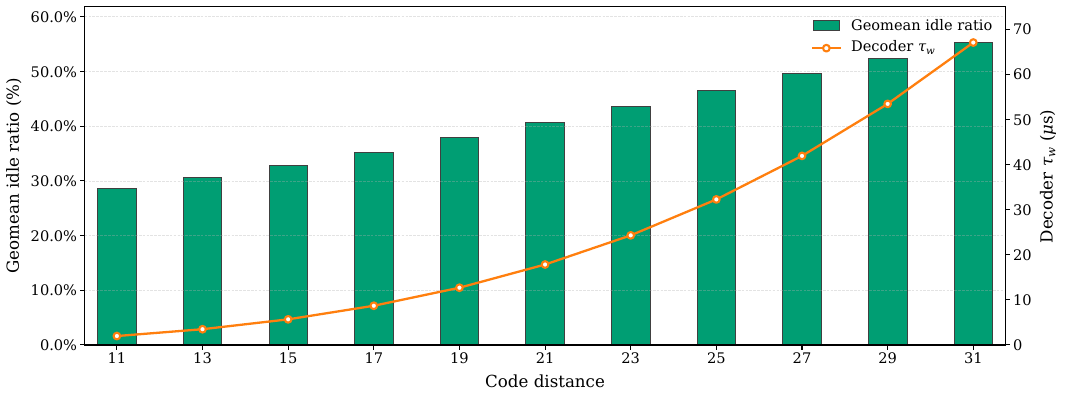}
    \caption{Geometric-mean decoding-induced idle-time fraction and decoder-window latency \(\tau_w\) across code distances. The idle-time fraction exceeds \(40\%\) at larger code distances.}
    \label{fig:execution-vs-idle}
\end{figure}

Figure~\ref{fig:execution-vs-idle} shows that this dependency has a substantial program-level cost. Across the evaluated workloads, decoding-induced stalls account for more than $40\%$ of execution time at larger code distances. Thus, even when parallel window decoding provides sufficient throughput to avoid a decoding backlog, waiting for \(T\)-gate decoder feedback remains a major contributor to program latency.\par

\begin{figure}
    \centering
    \includegraphics[width=0.9\linewidth]{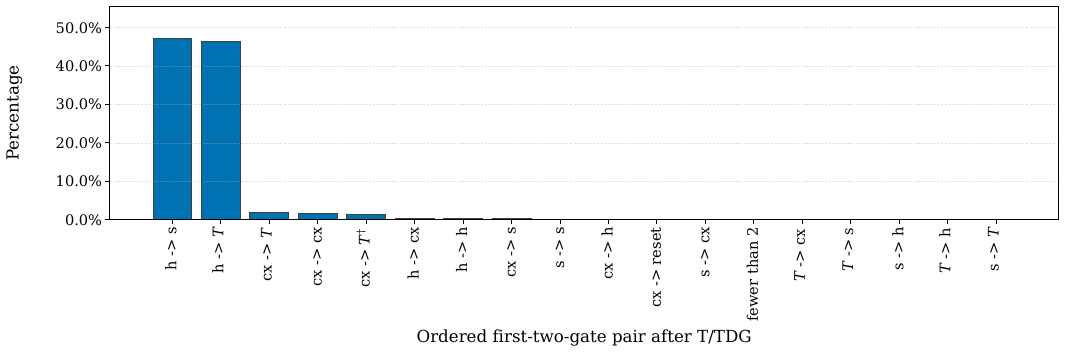}
    \caption{Distribution of the first two non-Pauli operations following $T/T^\dagger$ gates on the critical path.}
    \label{fig:dist-pair}
\end{figure}

This motivates \qsink{}: rather than only making decoder feedback available earlier, we can delay when that feedback becomes necessary by executing compatible operations following the $T$ gate. To quantify this opportunity, we transpile QASMBench circuits~\cite{li2023qasmbench} to Clifford+$T$ using gridsynth~\cite{ross2016optimal-gridsynth} and characterize the first two non-Pauli operations following each $T/T^\dagger$ on the critical path. As shown in Figure~\ref{fig:dist-pair}, the first post-$T$ operation is a Clifford gate in up to $98\%$ of the observed contexts, providing opportunities to execute compatible post-\(T\) operations before the conditional correction is resolved.

\section{Method}
\label{sec:method}

\begin{figure}[t]
    \centering
    \begin{subfigure}[t]{0.48\columnwidth}
        \centering
        \includegraphics[width=\linewidth]{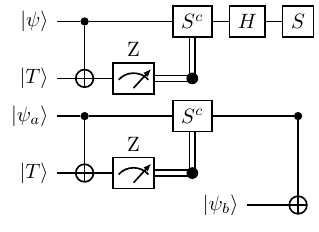}
        \caption{}
        \label{fig:first}
    \end{subfigure}
    \hfill
    \begin{subfigure}[t]{0.48\columnwidth}
        \centering
        \includegraphics[width=\linewidth]{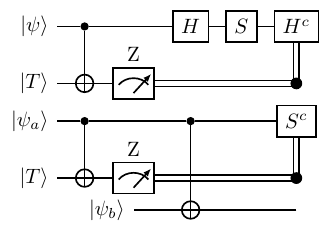}
        \caption{}
        \label{fig:second}
    \end{subfigure}
\caption{Representative \qsink{} transformations. The top shows the temporal sequence $T\rightarrow S_{\mathrm{cond}}\rightarrow H\rightarrow S$ transformed into $T\rightarrow H\rightarrow S\rightarrow H_{\mathrm{cond}}$, up to a Pauli-frame update. The bottom illustrates propagation through operations whose implementation begins with a $Z$-basis measurement, such as a subsequent $T/T^\dagger$ gate or a control-side measurement-based CNOT, allowing the pending conditional correction to be moved past the initial $M_Z$ step.}

    \label{fig:combined}
\end{figure}

We formalize here the \qsink{} transformations. Inspired by instruction sinking in classical compilers~\cite{gupta1997path}, \qsink{} moves a pending conditional $S/S^\dagger$ correction across compatible subsequent Clifford+$T$ operations while preserving program semantics. Figure~\ref{fig:combined} illustrates the resulting transformations for the representative \(SHS\) and \(T\)/CNOT cases. The following subsections derive the rewrite rules that define the transformation and establish its correctness. Throughout this section, temporal circuit sequences are written from left to right using arrows, while operator products follow the standard mathematical convention, with the rightmost operator acting first.

\subsection{S gate}
A pending conditional \(S\) correction commutes through any \(S\) or \(S^\dagger\) gate, since all are diagonal \(Z\)-axis Clifford rotations:
\[
\boxed{
S^b S^a = S^a S^b,\qquad a,b\in\{0,\pm1\}.
}
\]
Therefore, the pending correction commutes through an $S/S^\dagger$ gate unchanged.

\subsection{$H$ followed by $S/S^\dagger$}
Propagating \(S\) through \(H\) alone produces a non-Pauli Clifford correction that cannot be handled by a Pauli frame. We therefore target the common case where \(H\) is followed by \(S/S^\dagger\), which occurs in about \(50\%\) of the observed post-\(T\) patterns in Figure~\ref{fig:dist-pair}. Let \(S^m\), \(m\in\{0,1\}\), denote the conditional correction and \(S^b\), \(b\in\{+1,-1\}\), denote \(S\) or \(S^\dagger\). Moving the correction past the deterministic \(H\rightarrow S^b\) block gives

$$
S^bHS^mT=C_m^{(b)}S^bHT,
\qquad
C_m^{(b)}=S^bHS^mHS^{-b}.
$$

For \(m=0\), \(C_0^{(b)}=I\), while for \(m=1\),

$$
C_1^{(+)}=e^{i\pi/4}XH,
\qquad
C_1^{(-)}=e^{i\pi/4}ZH.
$$

Hence, up to global phase,

$$
\boxed{
C_0^{(+)}=C_0^{(-)}=I,\qquad
C_1^{(+)}\equiv XH,\qquad
C_1^{(-)}\equiv ZH.
}
$$

Thus, delaying the conditional \(S\) through either \(H\rightarrow S\) or \(H\rightarrow S^\dagger\) leaves only an \(H\) as the physical correction; the accompanying \(X\) or \(Z\) is absorbed into the Pauli frame. Since this replaces a conditional \(S\) of duration \(d\) with a conditional \(H\) of duration \(3d\), the rewrite is profitable only if

$$
\tau_D+2\tau_w>2d.
$$

Accordingly, \texttt{qSink-compiler} applies the transformation only when this condition is satisfied.


\subsection{T gate}
\label{sec:t-gate}

A $T$-gate teleportation introduces a measurement-conditioned $S$ correction, while a $T^\dagger$ introduces a measurement conditioned $S^\dagger$ correction. We write both cases as

$$
S^{\sigma a},
\qquad
a\in\{0,1\},\quad
\sigma\in\{+1,-1\},
$$

where \(\sigma=+1\) corresponds to \(S^a\) and \(\sigma=-1\) to \(S^{-a}\).

Both \(S\) and \(S^\dagger\) are diagonal in the \(Z\) basis and therefore commute with a \(Z\)-basis measurement. For

$$
P_m=\frac{I+(-1)^m Z}{2},
$$

we have

$$
[P_m,S^{\sigma a}]=0,
$$

so

$$
\|P_mS^{\sigma a}|\psi\rangle\|^2
=
\|P_m|\psi\rangle\|^2.
$$

Thus, the conditional correction does not affect the outcome probabilities of a subsequent \(Z\)-measurement.

Moreover, if the qubit is discarded after measurement,

$$
S^{\sigma a}|m\rangle
=
i^{\sigma a m}|m\rangle,
$$

so the remaining effect is only a branch-dependent global phase. Hence, for a measured-and-discarded qubit:

$$
\boxed{
M_Z S^{\sigma a} \equiv M_Z
}
$$

Consequently, the initial $Z$-basis measurement of a subsequent $T/T^\dagger$ gadget can be performed before the pending conditional $S/S^\dagger$ correction is resolved. \qsink{} exploits this property to move the decoder-dependent correction past the initial $M_Z$ step, postponing the point at which the correction must be resolved.

\subsection{CNOT gate}
The measurement-based CNOT is already compiled with the $Z$-type measurement before the subsequent $X$-type measurement. Since this $Z$-measurement consumes the control qubit, by Section~\ref{sec:t-gate}, a delayed \(S_q^a\) correction on the control does not need to be resolved before the measurement:

$$
M_{Z,q} S_q^a \equiv M_{Z,q}.
$$

Thus, \qsink{} can move a pending control-side correction past the initial $M_Z$ step, preventing it from blocking the beginning of the CNOT gadget.

\subsection{Extension to $S^\dagger$ and $T^\dagger$}
For a \(T^\dagger\) gate, the feed-forward correction is \(S^\dagger\) instead of \(S\); equivalently, all rewrite rules are obtained by replacing \(T\rightarrow T^\dagger\) and \(S\rightarrow S^\dagger\). The optimization is unchanged: the delayed correction commutes through $S/S^\dagger$ and the relevant $Z$-basis measurements, while the $H\rightarrow S/S^\dagger$ transformation produces the same conditional $H$ correction up to a Pauli-frame update.

\section{Implementation}
\begin{figure}
    \centering
    \includegraphics[width=0.8\linewidth]{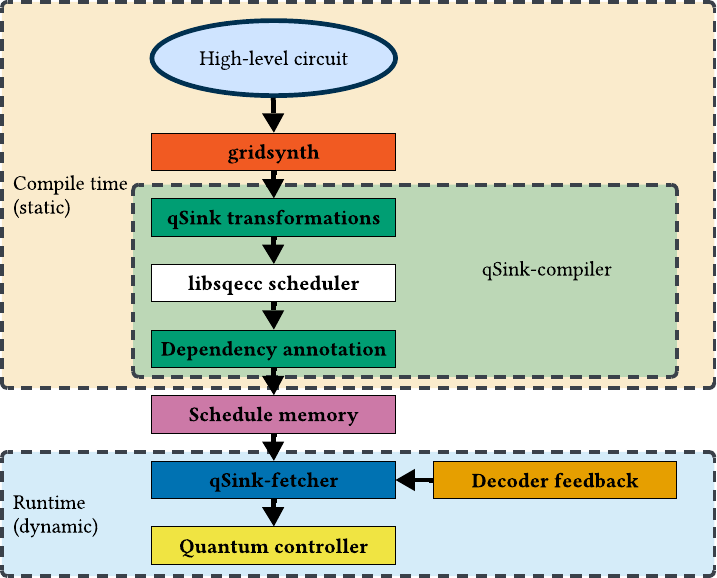}
    \caption{Compiler and runtime flow. \qsinkc{} applies correction-sinking transformations, uses the \texttt{liblsqecc} scheduler to generate the schedule, and annotates decoder-dependent corrections. At runtime, \qsinkf{} uses decoder feedback to retain or remove these corrections before forwarding the updated schedule to the controller.}
    \label{fig:flow}
\end{figure}

Figure~\ref{fig:flow} summarizes the \qsink{} compiler-runtime flow. We implement \qsinkc{} as a wrapper around the static Lattice Surgery Compiler~(\texttt{liblsqecc}~\cite{watkins2024high-liblsqecc}), retaining its scheduler as an unmodified backend. \qsinkc{} applies the transformations from Section~\ref{sec:method} and annotates decoder dependent corrections in the generated schedule with their measurement dependencies. At runtime, \qsinkf{} uses these annotations together with decoder feedback to update the schedule before forwarding it to the quantum controller.
\subsection{Circuit Transpilation}
Each circuit is lowered to Clifford+$R_Z$, with the remaining $R_Z$ rotations transpiled into Clifford+$T$ sequences using \texttt{gridsynth}~\cite{ross2016optimal-gridsynth}. Operations already supported by \texttt{liblsqecc} are preserved, and the resulting \texttt{QASM} is passed to its lattice-surgery compilation flow.

\subsection{\qsinkc{}}
\qsinkc{} first lowers multi-step logical operations, including $CNOT$ and $T/T^\dagger$, into lattice-surgery primitives. It then identifies post-$T$ contexts matching the rewrite rules in Section~\ref{sec:method} and moves compatible operations before the pending conditional correction. Based on the code distance, \qsinkc{} selects the set of transformations that minimizes schedule latency. \par
We extend the intermediate gate set with conditional $S$, conditional $H$, $M_X$, and $M_Z$ operations to represent these transformations explicitly. The rewritten program is then scheduled by the unmodified \texttt{liblsqecc} backend. Finally, \qsinkc{} annotates each decoder-dependent correction with the measurement outcome controlling it, preserving the dependency after operations are moved in the schedule and allowing the runtime to resolve the correction without recompilation or rescheduling.

\subsection{\qsinkf{}}

\qsinkf{} is a lightweight FPGA schedule preprocessor placed between the decoder, schedule memory, and quantum controller. Conditional corrections remain in their statically generated time slices and are tagged by \qsinkc{} with their decoder dependencies. Once the corresponding decoder result becomes available, \qsinkf{} uses this metadata to determine whether each conditional correction should be retained or removed, and forwards the updated time slice to the quantum controller.

Because all scheduling decisions are made statically, \qsinkf{} performs only decoder-dependent filtering. The main controller can therefore continue to consume a linear stream of time slices, while the existing compiler scheduler and inner decoder remain unchanged.

\section{Evaluation}
We evaluate four configurations: \texttt{liblsqecc} with and without \swiper{}~\cite{viszlai2025swiper}, and \qsinkc{} with and without \swiper{}. We model speculative window decoding using the latency model described in Section~\ref{sec:bg}, assuming $90\%$ correct speculation and the same code-distance-dependent decoder latencies used throughout our evaluation. We first compare the two techniques independently to determine whether compiler-level scheduling provides benefits comparable to decoder-level speculation. We then combine them to quantify their cumulative benefit. Finally, we quantify the hardware overhead of \qsinkf{}, which resolves decoder-dependent corrections at runtime.

\subsection{Experimental Setup}
Our evaluation uses QASMBench circuits~\cite{li2023qasmbench}, spanning multiple algorithmic classes and qubit counts. Table~\ref{tab:benchmark-circuits} summarizes the selected workloads and their post-decomposition characteristics. We exclude circuits with no $T$ gates, classically conditioned operations unsupported by \texttt{liblsqecc}~\cite{watkins2024high-liblsqecc}, or compilation memory requirements exceeding the 1536~GB available on a single compute node.
\begin{table*}[t]
\centering
\caption{Benchmark circuit classes, qubit counts, circuit depths, and $T$-path counts after Clifford+$T$ decomposition, ordered by increasing maximum $N_T$. For circuit classes with multiple instances, entries report the minimum--maximum range across the available instances. $N_T$ denotes the maximum number of $T$-gates on any qubit path, and $T/\mathrm{LP}$ is the percentage of circuit depth occupied by $T$ gates on a longest-depth path.}
\label{tab:benchmark-circuits}
\scriptsize
\setlength{\tabcolsep}{1.5pt}
\renewcommand{\arraystretch}{0.92}

\begin{minipage}[t]{0.495\textwidth}
\centering
\begin{tabular}{@{}>{\raggedright\arraybackslash}p{0.2\linewidth}
                    >{\raggedright\arraybackslash}p{0.3\linewidth}ccc@{}}
\toprule
\textbf{Circuit} & \textbf{Logical Qubits} & \textbf{Depth} & $N_T$ & $T/\mathrm{LP}$ \\
\midrule
QEC encoding & 5 & 17 & 1 & 5.9\% \\
Teleportation & 3 & 6 & 1 & 16.7\% \\
Bell & 4 & 135 & 2 & 1.5\% \\
Fredkin & 3 & 11 & 4 & 36.4\% \\
Toffoli & 3 & 10 & 4 & 40.0\% \\
Multiply & 13 & 19 & 6 & 31.6\% \\
Adder & 4, 10, 28, 64, 118, 433 & 11--37 & 2--10 & 18.2--27.0\% \\
Big adder & 18 & 37 & 10 & 27.0\% \\
SECA & 11 & 61 & 16 & 26.2\% \\
QRAM & 20 & 80 & 32 & 40.0\% \\
SAT & 7, 11 & 55--256 & 14--96 & 25.5--37.5\% \\
KNN & 25, 31, 41, 67, 129, 341 & 287--852 & 106--170 & 19.9--36.3\% \\
SWAP test & 25, 41, 83, 115, 361 & 291--902 & 105--180 & 20.0--36.6\% \\
Linear solver & 3 & 539 & 198 & 36.7\% \\
W state & 3, 27, 36, 76, 118 & 273--588 & 100--218 & 36.6--37.5\% \\
Multiplier & 15, 45, 75 & 153--765 & 54--270 & 35.3\% \\
QUGAN & 39, 71, 111 & 1148--1167 & 408--411 & 34.6--35.3\% \\
\bottomrule
\end{tabular}
\end{minipage}
\hfill
\begin{minipage}[t]{0.495\textwidth}
\centering
\begin{tabular}{@{}>{\raggedright\arraybackslash}p{0.2\linewidth}
                    >{\raggedright\arraybackslash}p{0.3\linewidth}ccc@{}}
\toprule
\textbf{Circuit} & \textbf{Logical Qubits} & \textbf{Depth} & $N_T$ & $T/\mathrm{LP}$ \\
\midrule
Basis test & 4 & 1307 & 502 & 38.4\% \\
QPE & 9 & 2015 & 794 & 39.4\% \\
VQE & 4 & 2091 & 799 & 38.0\% \\
Basis change & 3 & 3032 & 1100 & 36.3\% \\
QF21 & 15 & 4019 & 1590 & 39.6\% \\
QAOA & 3, 6 & 769--6258 & 300--2284 & 36.5--39.0\% \\
Ising & 10, 26, 34, 42, 66, 98, 420 & 1034--7136 & 407--2796 & 38.5--39.4\% \\
PEA & 5 & 7557 & 2940 & 38.9\% \\
GCM h6 & 13 & 12104 & 4338 & 35.8\% \\
DNN & 2, 8, 16, 33, 51 & 2162--12691 & 808--4398 & 34.6--37.4\% \\
Random QAOA ($k=3$, $p=1$) & 1000 & 15537 & 6082 & 39.1\% \\
QFT & 4, 18, 29, 63, 160 & 1029--25048 & 399--9768 & 38.8--39.7\% \\
Basis trotter & 4 & 29440 & 10414 & 35.4\% \\
Square root & 18, 45, 60 & 390--48270 & 130--16090 & 33.3\% \\
VQE--UCCSD & 6, 8, 28 & 22843--290513 & 8815--103202 & 35.5--38.6\% \\
HHL & 7, 10 & 64421--5880934 & 24421--2213264 & 37.6--37.9\% \\
\bottomrule
\end{tabular}
\end{minipage}
\end{table*}

Each circuit is transpiled to Clifford+$T$ using \texttt{gridsynth}~\cite{ross2016optimal-gridsynth} and compiled with both \texttt{liblsqecc} and \qsinkc{}. Compilation experiments run on AMD EPYC 9354 Zen 4 ``Genoa'' nodes with 64 cores and 1536~GB of RAM.\par
We evaluate code distances $d=11$ to $d=31$, chosen to closely match the range of prior speculative window decoding work used for comparison~\cite{viszlai2025swiper} and to cover a representative range for practical fault-tolerant quantum algorithms~\cite{alexeev2025artificial}.
\par
We implement \qsinkf{} on a Xilinx Alveo U250 FPGA (\texttt{xcu250-figd2104-2L-e}) using Vivado 2023.1. This implementation quantifies the hardware overhead of consuming decoder feedback and retaining or removing decoder-dependent corrections from the schedule at runtime.

\subsection{Results}
\subsubsection{Idle-Time Reduction}
\begin{figure}
    \centering
    \includegraphics[width=0.8\linewidth]{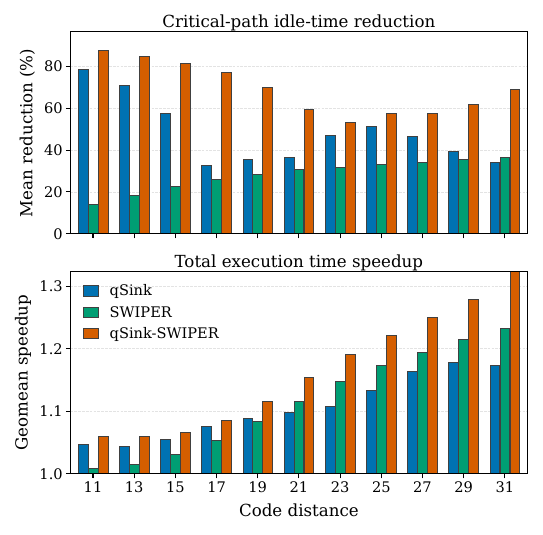}
    \caption{Critical-path decoding-induced idle-time reduction~(reported as arithmetic mean) and end-to-end speedup relative to baseline \texttt{liblsqecc} across code distances for \qsink{}, \swiper{}~(SWIPER), and their combination.}

    \label{fig:performance}
\end{figure}

Figure~\ref{fig:performance} reports the reduction in decoding-induced idle time on the critical path across code distances. \qsink{} removes a large fraction of this idle time at lower code distances, where compatible post-$T$ operations can overlap most of the wait for decoder feedback. Its effectiveness varies non-monotonically with code distance: around $d=19$, the $H\rightarrow S/S^\dagger$ transformation becomes profitable, increasing the available latency-hiding opportunities. At larger distances, however, decoder latency increasingly dominates gate-execution time, reducing the fraction of the stall that \qsink{} can hide while making \swiper{} increasingly effective.

Combining \qsink{} and \swiper{} consistently provides the largest reduction. The improvements are not additive because both techniques act on the same feedback interval shown in Figure~\ref{fig:t-gate}: \qsink{} overlaps execution with $\tau_D$, $\tau_{w1}$, and $\tau_{w2}$ by delaying when decoder feedback is required, while \swiper{} reduces the decoder-side contribution by speculatively resolving the dependency during $\tau_{w1}$ and, when necessary, $\tau_{w2}$.

\subsubsection{End-to-End Speedup}
The reduction in decoding-induced idle time yields end-to-end speedup, as shown in Figure~\ref{fig:performance}, but the two metrics do not map one-to-one. In particular, the $H\rightarrow S/S^\dagger$ transformation replaces a conditional $S$, of duration $d$, with a conditional $H$, of duration $3d$. It can therefore reduce decoder-induced idle time while introducing $2d$ of gate-execution latency, so a larger idle-time reduction does not imply a proportionally larger reduction in total execution time.

Within the evaluated range, the relative benefit of \qsink{} and \swiper{} is governed by their different scaling. The latency that \qsink{} can hide is determined by the duration of compatible post-$T$ operations, which scales with code distance $d$, whereas \swiper{} targets the decoder-window latency $\tau_w$, which grows faster than linearly with $d$ in our decoder model. Consequently, \qsink{} is more effective while gate-execution latency remains comparable to the decoder-feedback stall, but \swiper{} eventually surpasses standalone \qsink{} as $\tau_w$ grows more rapidly. 
 Their combination consistently achieves the highest speedup, reaching up to \combspeedup{} at $d=31$.

\subsubsection{\qsinkf}
\begin{figure}
    \centering
    \includegraphics[width=0.8\linewidth]{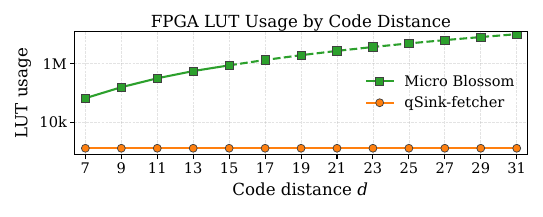}
    \caption{Comparison of FPGA resource utilization of the QEC decoder~\cite{wu2025micro} and \qsinkf{} within the system, with decoder utilization extrapolated to d=31 (dashed lines).}
    \label{fig:fpga_res_compare}
\end{figure}

Figure~\ref{fig:fpga_res_compare} compares the FPGA resource utilization of \qsinkf{} with the Micro Blossom Decoder~\cite{wu2025micro}. \qsinkf{} uses at most $1.97\%$ of the LUTs required by Micro Blossom decoder~\cite{wu2025micro} across the evaluated code distances, showing that decoder-dependent schedule filtering can be supported with minimal hardware.

\section{Related work}

Prior work on real-time QEC has largely focused on producing decoding results faster. Parallel window decoding removes the exponential backlog of sequential decoding~\cite{skoric2023parallel}, while algorithmic and hardware accelerators reduce inner-decoder latency~\cite{barber2025real-synthetic-decoder-data,alavisamani2024promatch,wu2025micro}. These approaches improve throughput and decoder-feedback latency, but execution still stalls once a feed-forward result becomes immediately required. \texttt{qSink} instead restructures the logical schedule to execute compatible post-$T$ operations while decoding remains in progress, without modifying the underlying decoder implementation. \par

Speculative window decoding reduces decoder-feedback latency by speculating on boundary information between adjacent decoding windows~\cite{viszlai2025swiper}. \texttt{qSink} acts on the opposite side of the dependency: \swiper{} produces the result earlier, whereas \texttt{qSink} delays when the result becomes critical by sinking the conditional correction across compatible operations. The two techniques are therefore complementary, as confirmed by their combined speedup. \par

Other work addresses decoder provisioning, scheduling, and runtime organization~\cite{maurya2026case-elastic-decoder,kim2025constraint-decoder-allocation,chen2026triage-adaptive-parallel-window-scheduler,wu2024lego,zhang2026learning-nn-for-parallel-decoding}. These techniques decide how decoding resources are allocated or structured. \texttt{qSink} instead extends the effective deadline of blocking results, creating additional scheduling slack. To the best of our knowledge, \qsink{} is the first compiler-runtime approach to systematically exploit post-\(T\) circuit structure to overlap lattice-surgery execution with QEC decoding.

\section{Conclusion}

Parallel window decoding addresses decoder throughput, but dependencies between decoding windows can still delay the feedback required by feed-forward-dependent $T$ gates. \qsink{} reduces this waiting time by moving conditional corrections past compatible post-$T$ operations, eliminating up to \idletime{} of decoding-induced critical-path idle time across our workloads.

Within the practically relevant code-distance range evaluated in this work, the effectiveness of \qsink{} and speculative window decoding reflects their different scaling. \qsink{} exploits the latency of compatible post-$T$ operations, while speculative window decoding makes decoder feedback available earlier. Their combination consistently provides the largest benefit, reaching approximately \combspeedup{} end-to-end speedup at $d=31$. Finally, our FPGA implementation shows that \qsinkf{} uses at most $1.97\%$ of the LUT resources of the reference QEC decoder. Overall, these results establish compiler-level scheduling as a complementary mechanism for reducing the critical-path impact of real-time QEC.

\section{Acknowledgment}
We acknowledge support from SSF~(grants FUS21-0063, SIP21-0087). MG is supported by the KAW through WACQT.
The authors used ChatGPT and Codex for assistance with mathematical consistency checks, code development, and language refinement. The authors reviewed all AI-assisted outputs and take full responsibility for the final content.

\bibliographystyle{plain} 
\bibliography{refs} 

\end{document}